\documentclass{iaus}
\usepackage{graphicx}

\title[Spitzer survey of massive sfr] 
{A Spitzer/IRAC Survey of Massive Star-Forming Regions}

\author[L. E. Allen et al.]   
{Lori E. Allen$^1$, 
 Joseph L. Hora$^1$, S. Thomas Megeath $^1$, \break Lynne K. Deutsch$^{1}$, 
G. G. Fazio$^1$, Luis Chavarria$^{1,2}$, \break \and Rebecca W. Dell$^{1,3}$} 

\affiliation{$^1$Harvard-Smithsonian Center for Astrophysics, 
60 Garden Street, Cambridge, MA, 02138 USA\break email: lallen@cfa.harvard.edu\\[\affilskip]
$^2$University of Chile \\[\affilskip] 
$^3$Harvard University}

\pubyear{2005}
\volume{227}  
\pagerange{1--6}
\date{?? and in revised form ??}
\jname{Proceedings Title IAU Symposium}
\editors{A.C. Editor, B.D. Editor \& C.E. Editor, eds.}

\def\msun{M$_{\odot}$ }

\def\hii{H\,{\sc ii} }

\begin{document}

\maketitle

\begin{abstract}
We are conducting a survey of several regions of high-mass
star formation to assess their content and structure.
The observations include Spitzer observations, ground-based optical and
near-IR imaging surveys, and optical and IR spectra of objects and
locations in the molecular clouds.  The goal of the survey is to gain a
better understanding of the processes involved in high mass star formation
by determining the characteristics of the stars detected in these regions
and investigating the properties of the interstellar medium (ISM)
environment in which these stars form.  
In this contribution, we present results on the identification and spatial 
analysis of young stars in three 
clusters, W5/AFGL~4029, S255, and S235. First we show how the IRAC data are used 
to roughly segregate young stars according to their mid-infrared colors, into 
two groups corresponding the SED Class I and Class II young stellar objects. 
Then using the IRAC data in combination with 2MASS, we show how more young stars 
can be identified. Finally, we examine the spatial distributions of young stars 
in these clusters and find a range of morphologies and of peak surface densities. 


\end{abstract}

\firstsection 

\section{Introduction}

Star formation is a self regulating process: once stars form they
immediately begin to alter and disrupt their natal environments and
eventually destroy their parental molecular clouds. A detailed
understanding of this feedback is a necessary step toward developing a
theory of the star formation efficiency and initial mass function in
molecular clouds.
The destruction of giant molecular clouds by OB stars is best studied
in bright rimmed clouds, regions where an edge--on molecular cloud
surface is externally illuminated by nearby young massive stars.  The edge--on geometry of
the resulting molecular cloud/H\,{\sc ii} region interface provides a
cross section of the photoevaporation process.
At such an interface, the ionization front advances into
the molecular cloud as the H\,{\sc ii} region expands.
The shock fronts associated with these disruptive processes
may trigger secondary
or sequential star formation.
In order to investigate these processes, we have 
initiated a survey of massive star forming 
regions with prominent infrared photodissociation regions. In this contribution 
we report on the spatial distributions of young stars in three such regions. 

\section{Observations}

The Spitzer data were obtained with the InfraRed Array Camera (IRAC;
Fazio et al. 2004). 
For each region, a map was made in all 4 IRAC bands covering $\sim$ 0.25 deg$^2$. 
An integration time of 12 seconds per dither, with 3 dithers per map position, was
used.  Short integration frames (0.6 sec) were also obtained at each position, for
the recovery of bright sources that are saturated in the longer frames.  The Basic
Calibrated Data frames, products of the Spitzer Science Center pipeline, were used 
to construct mosaics in each IRAC band.  
Photometry was performed on the mosaics, 
and photometric completeness limits were estimated to be
16, 15, 14 and 13 mag. at 3.6, 4.5, 5.8 and 8 $\mu$m, respectively. 
 
\begin{figure}
  \includegraphics[height=3in,width=5in,height=2.5in]{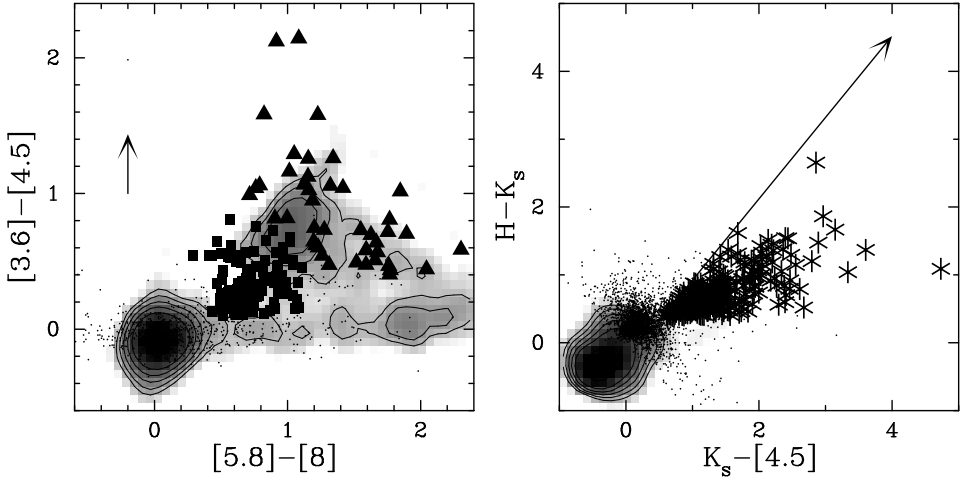}
\caption{{\bf Left:} IRAC color-color diagram for W5/AFGL~4029. Squares have 
Class II 
colors, triangles Class I, according to comparison with models of young stars.
Contours show the distribution of 
possible contaminating sources and start at N=5 (see text).
{\bf Right:} 2MASS 
H and K$_s$ bands are combined with IRAC 4.5 $\mu$m to select a greater number of 
stars with infrared excess, including those not detected in the longer-wavelength 
bands of IRAC. Contours are as at left.}
\end{figure}

\section{Identifying Young Stars} 

\subsection{IRAC only} 

By comparing the measured IRAC colors with model predictions, we have found 
that young stars occupy somewhat distinguishable regions in the IRAC color-color 
diagram, corresponding to their SED class (Allen et al. 2004; Megeath et al. 2004; 
see also Whitney et al. 2004). 
This was recently verified using 
data for the Taurus molecular cloud, in which the nature of individual  
sources is well known (Hartmann et al. 2005). 
The IRAC color-color diagram for W5/AFGL~4029 is shown in Figure 1 (left panel). 
Most of the sources are concentrated in 
three groups. Near (0,0) the sample consists of field stars, background stars with 
modest or no reddening, and young stars with no infrared excess emission in the 
IRAC bands (points). Since we cannot distinguish between these, we shall limit 
our analysis to the remaining two groups:  
one group occupying the range (x,y)=(0.4,0.2) to (1.1,0.8) corresponds to the 
colors of Class II objects, or young stars with accretion disks (squares), and 
a more dispersed group 
redward of the Class II domain containing Class I sources, or stars with  
disks and 
infalling envelopes (triangles). 

Due to the sensitivity of the IRAC observations and the relative transparency 
of molecular clouds in the mid-infrared, contamination of our sample by background 
stars and galaxies is probable. Here we attempt to quantify that contamination, 
using data from the IRAC Shallow Survey, which imaged 8.5 deg$^2$ in Bo\"{o}tes 
(Eisenhardt et al. 2004). 
Because the Shallow Survey was significantly deeper than our observations, we 
first selected only those sources with magnitudes equal to or brighter than the 
completeness limits of our study. Next we scaled the Shallow Survey data by the 
area of our map. The resulting distribution of IRAC colors from the Shallow Survey 
are plotted as contours in number of sources per map area, in Figure 1. 
Among the Class I and II sources, the highest concentration of background sources 
is in the boundary between the Class I 
and Class II domains, where there may be as many as 20 background sources, mostly AGN 
(Stern et al. 2005). 
In Figure 2, we show the spatial distributions of Class I and Class II sources 
in W5/AFGL~4029. Most of the Class I sources are concentrated in the embedded 
clusters near the edge of the molecular cloud, while the Class II sources 
are more widely distributed and include stars located in the adjacent \hii  region. 
In Figures 3 and 4, we plot the distributions of Class I and II sources in S255 and 
S235, respectively. They show similar behavior to W5 in the relative spatial 
distributions of Class I and II sources. 

\begin{figure}
\includegraphics[width=5in]{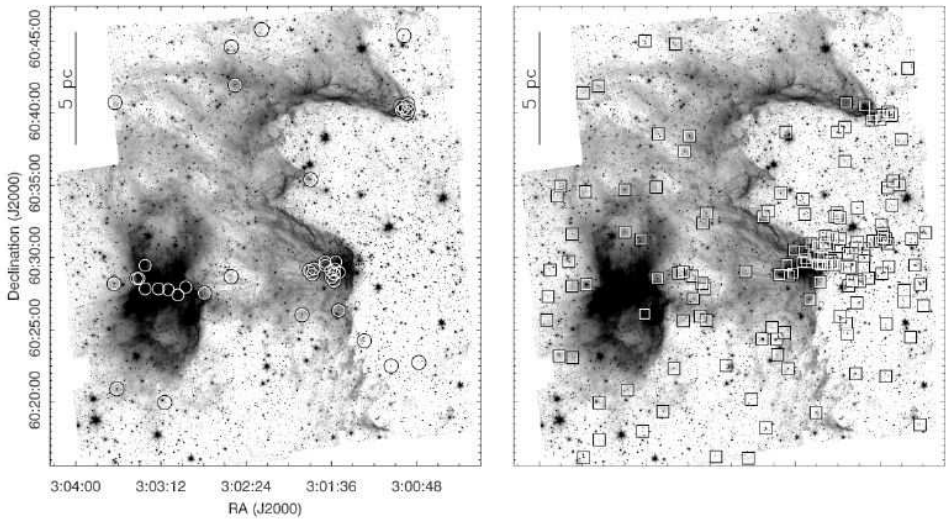}
\caption{{\bf Left:} Sources identified as Class I (triangles in Figure 1 left panel) 
are shown marked on the IRAC 4.5 $\mu$m mosaic of W5/AFGL~4029.  The Class I sources are 
clearly concentrated in the two embedded clusters near the edge of the molecular cloud, suggesting that these are currently the most active sites of star formation in the 
cloud. {\bf Right:} Class II sources (squares in Figure 1 left panel) are more widely 
distributed, and extend well off the molecular cloud into the adjacent \hii   region.} 
\end{figure}

\begin{figure}
\includegraphics[width=5in]{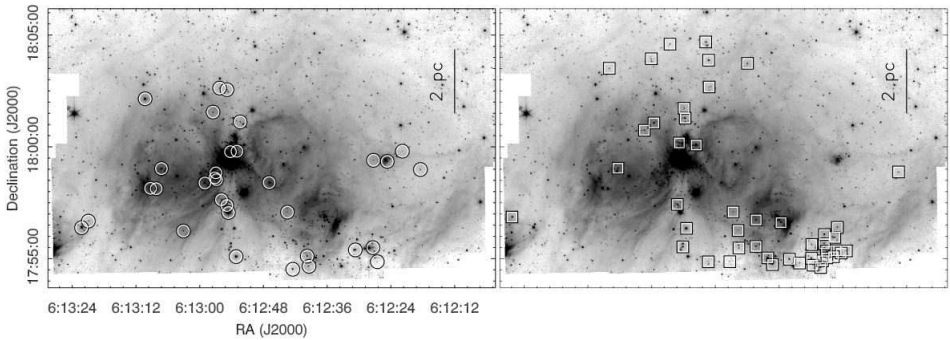}
\caption{{\bf Left:} Sources identified as Class I 
are shown marked on the IRAC 4.5 $\mu$m mosaic of S255. 
{\bf Right:} Class II sources are more widely 
distributed. }
\end{figure}

\begin{figure}
\includegraphics[width=5in]{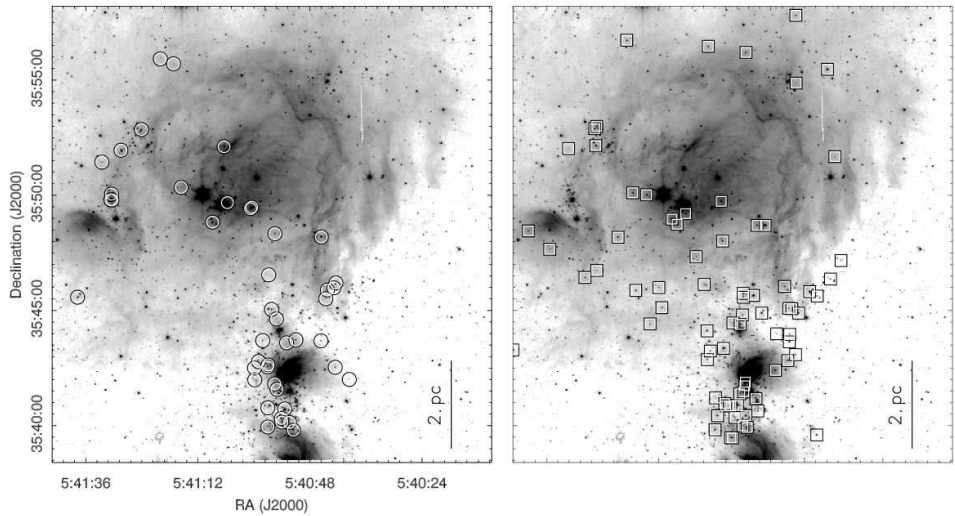}
\caption{Class I (left) and Class II (right) sources 
are shown marked on the IRAC 4.5 $\mu$m mosaic of S235.} 
\end{figure}

\subsection{IRAC and 2MASS combined}

While the IRAC colors are effective for finding Class I and Class II sources, 
the technique is limited to those stars detected in all four IRAC bands. 
Because IRAC is more sensitive to stellar photospheres at 3.6 and 4.5 $\mu$m than 
at 5.8 or 8.0 $\mu$m, we can use a combination of near-IR fluxes and 
fluxes in the first two IRAC bands to identify more young stars by their infrared 
excess. In Figure 1 (right panel) we show one such attempt, using 2MASS H, K$_s$ and IRAC 4.5 $\mu$m 
photometry. 
Points to the right of the reddening vector are stars with infrared excess. 
As in the left panel, we have plotted the contaminating background colors as 
contours of surface density. The numbers are negligible in the region of IR excess 
sources. 

\section{Spatial Distributions and Surface Densities of Young Stars} 

Using the combined 2MASS$+$IRAC data as described in the previous section, 
we plot the distribution of all IR excess sources in W5/AFGL~4029 (Figure 5, left panel). 
In general, they follow the distribution of the Class II sources in Figure 2,  
but are more numerous. In the right panel of Figure 5 is shown the surface density 
of the young stars, where the lowest contour level corresponds to approximately 
10\% of the peak value of $\sim$5000 stars pc$^{-2}$. 
At a distance of $\sim$2 kpc, W5 is comprised of several
overlapping bubbles of ionized gas, extending over 2 x 1.5 degrees on the sky.
The easternmost bubble is ionized by an O7 star. AFGL~4029 borders the eastern 
edge of this bubble, and has been the subject of several previous investigations.
Both Wilking  et al. (1984) and Karr \& Martin (2003)
analyzed the distribution of IRAS sources across the region.
Their consideration of the timescales for star formation and the expansion
of the H\,{\sc ii} region led them to conclude that triggering is a plausible
mechanism for the star formation observed there. 
The IRAC data support this scenario: the protostars are tightly clustered in 
two groups on the edge of
the molecular
cloud, coincident with the H\,{\sc ii}/molecular interface. 
The (presumably more evolved) Class II stars are more widely dispersed, both inside
the cloud and
in the adjacent H\,{\sc ii} region. 
The Spitzer/IRAC images also highlight
the interface between the molecular and ionized gas, and provide a dramatic illustration
of the edge--on geometry of the region. 
Spatial distributions and surface densities of S255 and S235 are shown in Figures 
6 and 7, respectively. 

\begin{figure}
\includegraphics[width=5in]{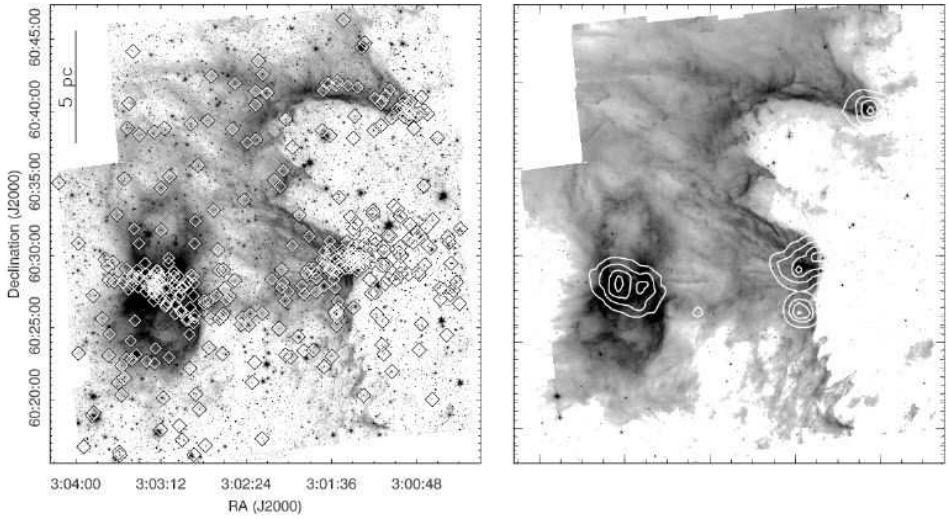}
\caption{{\bf Left:} The distribution of young stars as identified from the $H-K_s, K_s-4.5 \mu$m color-color 
diagram in Figure 1, plotted on the IRAC 4.5 $\mu$m mosaic. 
{\bf Right:} contours of the 
surface density of these sources, overlaid on the IRAC 8$\mu$m mosaic. 
Contour levels start at 10\% of the peak surface density and double at 
each interval.}
\end{figure} 

While the peak surface densities measured here range from $\sim$2000 stars pc$^{-2}$ 
in S235 to $\sim$5000 stars pc$^{-2}$ in W5/AFGL~4029 (with a peak in S255 of 
$\sim$4000 stars pc$^{-2}$),  
our survey is complete only to stars of spectral type $\sim$K5 for a population of 
age $\sim$2 Myr. 
Assuming these clusters are forming stars in accordance with the 
IMF measured in nearby star-forming regions (Meyer et al. 2000),  
then the {\it actual} surface densities are probably 2--10 times higher. 

\begin{figure}
\includegraphics[width=5in]{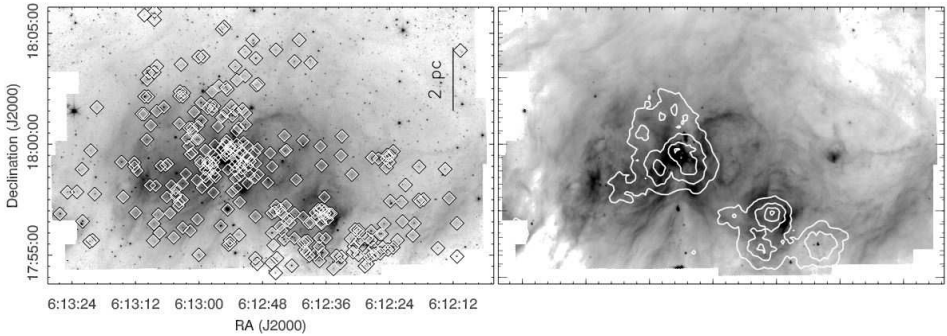}
\caption{{\bf Left:} The distribution of young stars as identified from a combination 
of 2MASS and IRAC 4.5 $\mu$m band in S255, and {\bf right:} contours of the 
surface density of these sources. Contour levels are as in Figure 5.}
\end{figure} 

\begin{figure}
\includegraphics[width=5in]{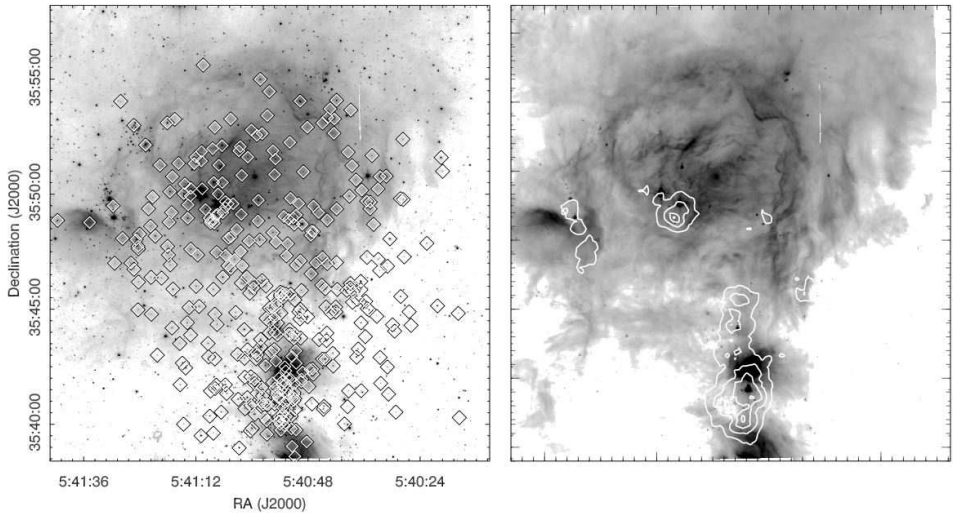}
\caption{{\bf Left:} The distribution of young stars as identified from a combination 
of 2MASS and IRAC 4.5 $\mu$m band in S235, and {\bf right:} contours of the 
surface density of these sources. Contour levels are as in Figure 5.}
\end{figure} 

\section{Summary} 

We have conducted a Spitzer/IRAC survey of the three massive star-forming regions 
W5/AFGL~4029, S255 and S235. We find that IRAC colors are useful for identifying 
Class I and II young stellar objects. When combined with near-IR fluxes from 
2MASS, all of the young stars with infrared excess can be identified. Class II 
objects are more widely distributed than Class I, and the IR-excess sources identified 
using IRAC$+$2MASS are more widely distributed still. The peak surface densities of 
young stars of mass $\ge$1 \msun  range from 2000 to 5000 stars pc$^{-2}$. When 
the incompleteness of the survey is taken into account, peak surface densities may be higher by a factor of 
2 to 10. 

\begin{acknowledgments}

The original P.I. of this IRAC GTO program was Lynne K. Deutsch, who 
died on April 2, 2004. She is dearly missed. This work is based on observations made with the Spitzer Space Telescope, 
operated by the Jet Propulsion Laboratory under NASA contract 1407.
 
\end{acknowledgments}

\end{document}